\documentclass[floatfix, preprint, onecolumn, prd,showpacs, showkeys, preprintnumbers, nofootinbib, superscriptaddress]{revtex4-1}
\usepackage{times}
\usepackage{amssymb,amsbsy,amsmath,amsfonts}
\usepackage{graphicx}
\usepackage{float}
\usepackage{color}
\usepackage{morefloats}
\usepackage{rotating}
\usepackage{srcltx}
\usepackage{slashed}
\usepackage{multirow}
\usepackage{verbatim}
\usepackage{hyperref}
\usepackage{tabularx}

\begin{document}

\title{$B_s\pi$--$B\bar{K}$ interactions in finite volume and the $X(5568)$ }

\author{Jun-Xu Lu}
\affiliation{School of Physics and
Nuclear Energy Engineering, International Research Center for Nuclei and Particles in the Cosmos,
Beijing Key Laboratory of Advanced Nuclear Materials and Physics,  Beihang University, Beijing 100191, China}

\author{Xiu-Lei Ren}
\affiliation{State Key Laboratory of Nuclear Physics and Technology, School of Physics, Peking University, Beijing 100871, China}

\author{Li-Sheng Geng}
\email[E-mail me at: ]{lisheng.geng@buaa.edu.cn}
\affiliation{School of Physics and
Nuclear Energy Engineering, International Research Center for Nuclei and Particles in the Cosmos,
Beijing Key Laboratory of Advanced Nuclear Materials and Physics,  Beihang University, Beijing 100191, China}

\begin{abstract}
The recent observation of $X(5568)$ by the D0 Collaboration has aroused a lot of interest both theoretically and experimentally.
In the present work, we first point out that $X(5568)$ and $D_{s0}^*(2317)$ cannot  simultaneously be of molecular nature, from the perspective of heavy-quark symmetry and chiral symmetry, based on a previous study of the lattice QCD scattering lengths of $DK$ and its coupled channels.  Then we compute the discrete energy levels of the  $B_s\pi$ and $B\bar{K}$ system in finite volume  using unitary chiral perturbation theory.
The comparison with the latest lattice QCD simulation, which disfavors the existence of $X(5568)$, supports our picture where the  $B_s\pi$ and $B\bar{K}$ interactions are weak and $X(5568)$ cannot be a $B_s\pi$ and $B\bar{K}$ molecular state. In addition, we show that the extended Weinberg compositeness condition also
indicates that $X(5568)$ cannot be a molecular state made from  $B_s\pi$ and $B\bar{K}$ interactions.

\end{abstract}


\date{\today}

\maketitle
\section{Introduction}
Recently,  an apparently exotic mesonic state, the so-called  $X(5568)$ state was observed by
the D0 Collaboration in the $B^0_s\pi^\pm$ invariant mass spectrum~\cite{D0:2016mwd}. The extracted mass and width are $M = 5567.8 \pm 2.9^{+0.9}_{-1.9}$ MeV and $\Gamma = 21.9\pm6.4^{+5.0}_{-2.5}$ MeV, respectively, and the preferred spin-parity  is $J^P = 0^+$.  This state, being one of the exotic $XYZ$ states~\cite{Chen:2016qju}, should contain at least four valence quark flavors $u$, $d$, $s$, and $b$.

The experimental observation of $X(5568)$ has inspired much theoretical work.  It has been proposed to be  either a tetraquark state~\cite{Tang:2016pcf,Chen:2016mqt,Stancu:2016sfd,Liu:2016ogz,Zanetti:2016wjn,Zanetti:2016wjn,Dias:2016dme,Wang:2016mee,Wang:2016wkj,Agaev:2016mjb,Agaev:2016urs,Agaev:2016ijz,Lu:2016zhe}, a triangular singularity~\cite{Liu:2016xly}, or  a molecular state~\cite{Xiao:2016mho}.
Although these  theoretical studies  favor the existence of a state that can be identified as the $X(5568)$,  some other studies have yielded negative conclusions. For instance, the difficulty to accommodate such a narrow structure with a relatively low mass  has been stressed by Burns \textit{et al.}~\cite{Burns:2016gvy} and Guo \textit{et al.}~\cite{Guo:2016nhb}. A recent study in the chiral quark model showed that neither diquark--antidiquark nor meson--meson structures  support the existence of $X(5568)$~\cite{Chen:2016npt}. In Ref.~\cite{Wang:2016tsi}, the lowest-lying tetraquark $s$-wave state was found to be 150 MeV higher than $X(5568)$.  On the experimental side,  neither the LHCb  nor the CMS Collaboration  found a signal  corresponding to $X(5568)$~\cite{lhcb,cms}.

Approximate heavy-quark symmetry and its breaking pattern provide a powerful tool to understand the nature of $X(5568)$. In the charm sector,
$D_{s0}^*(2317)$ has been suggested to be a molecular state made from $D K$ and $D_s\eta$ interactions in many studies; see, e.g., Refs.~\cite{Bardeen:2003kt,Kolomeitsev:2003ac,Guo:2006fu,Cleven:2010aw,Altenbuchinger:2013vwa,Liu:2012zya}.\footnote{In Refs.~\cite{Bardeen:2003kt,Kolomeitsev:2003ac,Guo:2006fu,Cleven:2010aw,Altenbuchinger:2013vwa}, the bottom parter of
$D_{s0}^*(2317)$ was also predicted.}  Naively, $X(5568)$ might be a heavy-quark partner of the $D_{s0}^*(2317)$. However, in  Ref.~\cite{Guo:2016nhb}, the authors point out that if one assumes  a molecular picture for the $X(5568)$, heavy-quark symmetry then dictates that the charmed partner of $X(5568)$ should be located around $2.24\pm0.15$ GeV. So far, no signal has been reported yet with the associated quantum numbers in this energy region~\cite{Guo:2016nhb}. Note that the QCD sum rules predicted that the charmed partner of $X(5568)$ is located at much higher energies about $2.5-2.6$ GeV~\cite{Agaev:2016lkl,Chen:2016mqt}.

The unitary chiral perturbation theory (UChPT), respecting both approximate heavy-quark symmetry and chiral symmetry and their breaking pattern, has turned out to be a useful tool in understanding the $XYZ$ states. In Ref.~\cite{Altenbuchinger:2013vwa}, using the lattice QCD (lQCD) scattering lengths~\cite{Liu:2012zya} to fix the relevant low energy constants (LECs) of the covariant UChPT, one found that the $D_{s0}^*(2317)$ emerges naturally, similar to Refs.~\cite{Liu:2012zya,Guo:2015dha,Yao:2015qia}. When
extended to the bottom sector guided by heavy-quark symmetry, the covariant UChPT shows that the interactions in
the $B_s\pi$--$B\bar{K}$ channel are rather weak and do not support the existence of a low-lying resonance or bound state, consistent with
 an explicit search on the second Riemann sheet~\cite{Altenbuchinger:2013vwa}.

Nevertheless, in Ref.~\cite{Albaladejo:2016eps}, the authors performed a fit to the D0 invariant mass distribution~\cite{D0:2016mwd} by employing the $B_s\pi$ and $B\bar{K}$ coupled channel unitary chiral amplitudes and treating the unknown subtraction constant as a free parameter.
The best fit yields  a dynamically generated state consistent with $X(5568)$~\cite{D0:2016mwd}.
Nevertheless, the authors noted that a large cutoff $\Lambda$, compared with a ``natural" size of about 1 GeV,  is needed to describe the D0 data. The unusual size of the cutoff points clearly to the presence of missing channels, contributions of other sources, or existence of ``non-molecular'' components, such as sizable tetraquark configurations, in the framework of the UChPT. As a result, it was concluded that a pure molecular state, dynamically generated by the unitary loops, is disfavored.   A similar conclusion was reached in a later study utilizing $p$-wave coupled channel dynamics in the UChPT~\cite{Kang:2016zmv}. It should be pointed out that irrespective of the nature of $X(5568)$, the UChPT of
Ref.~\cite{Albaladejo:2016eps} provides a good description of the D0  data. Therefore,  it needs to be further tested.

In the present work, we would like to formulate the UChPT of Ref.~\cite{Altenbuchinger:2013vwa,Albaladejo:2016eps} in  finite volume and compute relevant discrete energy levels and scattering lengths.
A comparison with  lattice QCD simulations then allows one to distinguish these two scenarios and provides more clues about the nature of  the $X(5568)$. In the remaining of this paper,  we denote  the UChPT of Ref.~\cite{Albaladejo:2016eps} by $X$-UChPT and that of Ref.~\cite{Altenbuchinger:2013vwa} by $\slashed{X}$-UChPT to indicate that one of them dynamically generates $X(5568)$ and the other not.

 This paper is organized as follows. In Sect. 2, we briefly describe the UChPT of Ref.~\cite{Altenbuchinger:2013vwa} and Ref.~\cite{Albaladejo:2016eps} and then in Sect. 3 we point out that  from the perspective of heavy-quark symmetry and chiral symmetry as implemented in UChPT, the lQCD scattering lengths of $DK$ and its coupled channels imply that the $B_s\pi$ and $B\bar{K}$ interactions are rather weak and do not support a molecular state that can be identified as $X(5568)$. As a result,  a typical molecular picture for $X(5568)$ similar to that for the $D_{s0}^*(2317)$ case is not favored in UChPT, which is supported by the extended Weinberg compositeness condition. In Sect. 4, we formulate
the UChPT of Ref.~\cite{Altenbuchinger:2013vwa,Albaladejo:2016eps} in a finite box and calculate the discrete energy levels that
can be extracted in a lattice QCD simulation.  The results are contrasted with the latest lQCD simulation of Ref.~\cite{Lang:2016jpk}, followed by a short summary in Sect. 5.
\section{Unitary chiral perturbation theory}
UChPT has two basic building blocks, a kernel potential provided by chiral perturbation theory and a unitarization procedure. The kernel potentials constrained by chiral symmetry and other
relevant symmetries, such as heavy-quark symmetry in the present case, are standard in most cases, while the unitarization procedures can differ in their treatment of left-hand cuts or higher order effects, although
they all satisfy two-body elastic unitarity.

The leading order kernel potential employed in Refs.~\cite{Altenbuchinger:2013vwa,Albaladejo:2016eps}  has the following form:
\begin{equation}\label{eq:LO}
V_{ij}=\frac{C_{ij}}{8f^2}\left( 3s-(M_i^2+m_i^2+M_j^2+m_j^2)-\frac{\Delta_1\Delta_2}{s}\right),
\end{equation}
where $i=1(2)$ denotes the $B_s\pi$ ($B\bar{K}$) channel, $s$ is the invariant mass squared of the system, $f$ is the pseudoscalar meson decay constant in the chiral limit,
$\Delta_i=M_i^2-m_i^2$ and $M_1(m_1)$ and $M_2(m_2)$ are the
masses of $B_s(\pi)$ and $B(\bar{K})$ mesons. The coefficients $C_{ij}$ are $C_{11}=C_{22}=0$, $C_{12}=C_{21}=1$.

In Refs.~\cite{Altenbuchinger:2013vwa,Albaladejo:2016eps}, the Bethe--Salpeter equation
is adopted to unitarize the chiral kernel obtained above. In the context of the UChPT,
the integral Bethe--Salpeter equation is often simplified and approximated as an algebraic equation with the use of the on-shell approximation.
\footnote{For a recent study of off-shell effects, see Ref.~\cite{Altenbuchinger:2013gaa} and the references cited therein.} It reads
\begin{equation}
T=V+VGT,
\end{equation}
where $T$ is the unitarized amplitude, $V$ the potential, and $G$ the one-loop 2-point scalar function.  In $n$ dimensions, $G$ has the following simple form:
\begin{equation}
   G_i=i\int\frac{\mathrm{d}^{n}q}{(2\pi )^{n}}\frac{1}{[(P-q)^{2}-m_i+i\epsilon][q^{2}-M_i^{2}+i\epsilon ]} \\
\end{equation}
where $P$ is the total center-of-mass momentum of the system.

The loop function $G$ is divergent and needs to be regularized. In the dimensional regularization scheme, it has the following form:
\begin{eqnarray}\label{Eq:MSbar}
   \begin{split}
      G_{\overline{\mathrm{MS}}}(s,M^{2},m^{2}) =& {}\frac{1}{16\pi^{2}}\left[\frac{m^{2}-M^{2}+s}{2s}\log\left(\frac{m^{2}}{M^{2}}\right)\right.\\
                                       & -\frac{q}{\sqrt{s}}(\log[2q\sqrt{s}+m^{2}-M^{2}-s]+\log[2q\sqrt{s}-m^{2}+M^{2}-s]\\
                                       & -\log[2q\sqrt{s}+m^{2}-M^{2}+s]-\log[2q\sqrt{s}-m^{2}+M^{2}+s])\\
                                       & \left.+\left(\log\left(\frac{M^{2}}{\mu^{2}}\right)+a\right)\right],
      \end{split}
\end{eqnarray}
where  $a$ is the subtraction constant, and $\mu$ the regularization scale and $s=P^2$.  The difference between Eq.~(\ref{Eq:MSbar}) and its counterpart in Ref.~\cite{Albaladejo:2016eps} is a constant and can be absorbed into the subtraction constant.

It has been noted that the relativistic loop function, Eq.~(\ref{Eq:MSbar}), violates heavy-quark symmetry and the naive chiral power counting. Several methods (see, e.g., Refs.~\cite{Kolomeitsev:2003ac,Gamermann:2006nm,Gamermann:2007fi,Cleven:2010aw}) have been proposed to deal with such a problem.
In   $\slashed{X}$-UChPT~\cite{Altenbuchinger:2013vwa}, the minimal subtraction scheme is modified to explicitly conserve  heavy quark symmetry and the naive chiral power counting. Confined to
either the charm sector or the bottom sector alone, the modified subtraction scheme, termed the heavy-quark-symmetry (HQS) inspired scheme, is equivalent to the $\overline{\mathrm{MS}}$ scheme, but
it can link both sectors in a way that conserves heavy-quark symmetry   up to order $1/M_H$, with $M_H$ the chiral limit  of the heavy hadron mass.\footnote{For an application
of the HQS scheme in the singly charmed (bottom) baryon sector, see Ref.~\cite{Lu:2014ina}.}
 The loop function of the HQS scheme is related with that of the $\overline{\mathrm{MS}}$ scheme via
\begin{equation}\label{Eq:HQS}
G_{\mathrm{HQS}}=G_{\overline{\mathrm{MS}}}-\frac{1}{16\pi^{2}}\left(\log\left(\frac{\mathring{M}^{2}}{\mu^{2}}\right)-2\right)+\frac{ m_\mathrm{sub}}{16\pi^{2}\mathring{M}}\left(\log\left(\frac{\mathring{M}^{2}}{\mu^{2}}\right)+a'\right),
\end{equation}
where $m_\mathrm{sub}$ is the average mass of the Goldstone bosons, $\mathring{M}$ the chiral limit value of the bottom (charm) meson masses and $a'$ the subtraction constant. In the
HQS scheme, the subtraction constant determined in the charm (bottom) sector is the same as that determined in the bottom (charm) sector, while this is not the case for the subtraction constant
of the $\overline{\mathrm{MS}}$ scheme. However, one may use the same cutoff in the cutoff scheme for both bottom and charm sectors related via heavy-quark symmetry
(for a different argument, see, e.g., Ref.~\cite{Ozpineci:2013qza}).

In Ref.~\cite{Altenbuchinger:2013vwa}, it was shown that the LO potential of Eq.~(\ref{eq:LO}) cannot describe the lQCD scattering lengths of Ref.~\cite{Liu:2012zya}. One needs go to the next-to-leading order (NLO).  At NLO, there are six  more LECs, namely $c_0$, $c_1$, $c_{24}$, $c_{35}$, $c_4$, $c_5$.  Among them,  $c_0$ is determined by fitting to the light quark mass dependence of lQCD $D$ and $D_s$ masses~\cite{Liu:2012zya}, and $c_1$ is determined by reproducing the experimental $D$ and $D_s$ mass difference. Once $c_0$ and $c_1$ are fixed, the remaining LECs and the subtraction constant are determined by fitting to the lQCD scattering lengths of Ref.~\cite{Liu:2012zya}, yielding a
$\chi^2/\mathrm{d.o.f.}=1.23$. With these LECs, $D_{s0}^*(2317)$ appears naturally  at $2317\pm10$ MeV. As a result, in the present work we employ the NLO UChPT of Ref.~\cite{Altenbuchinger:2013vwa}.

\section{Scattering lengths and compositeness }
\subsection{Scattering lengths}
The scattering lengths of a $D$ ($D_s$) meson with a Nambu--Goldstone pseudoscalar meson have been
studied on a lattice~\cite{Liu:2012zya}. With these scattering lengths as inputs, various
groups have predicted the existence of $D_{s0}^*(2317)$ and its counterparts both in the charm sector
and in the bottom sector~\cite{  Altenbuchinger:2013vwa,Guo:2015dha,Yao:2015qia}. It is worth pointing out that the predicted counterparts of the $D_{s0}^*(2317)$ and
$D_{s1}(2416)$
are indeed observed in a later lQCD simulation~\cite{Lang:2015hza}.  In the following, we  compare the scattering lengths of
$B_s\pi$, $B\bar{K}$, $D_s\pi$, and $DK$ obtained in $\slashed{X}$-UChPT~\cite{Altenbuchinger:2013vwa}
with those obtained in $X$-UChPT~\cite{ Albaladejo:2016eps}, to check the consistency between
the constraint imposed by the existence of the $X(5568)$ and that of $D_{s0}^*(2317)$ and the lQCD scattering lengths of Ref.~\cite{Liu:2012zya}.

\begin{table}
\centering
\caption{$B_s\pi$, $B\bar{K}$, $D_s\pi$, and $DK$ scattering lengths in $\slashed{X}$-UChPT~\cite{Altenbuchinger:2013vwa} and $X$-UChPT~\cite{Albaladejo:2016eps} \footnote{
For $X$-UChPT~\cite{Albaladejo:2016eps}, the scattering lengths are calculated using the loop function regularized in the cutoff scheme with the cutoff fixed by fitting to the D0 data. See Sect. 2 for
details.}, in units of fm. }\label{Table:twochannel}
\begin{tabular}{c|cc|c|cc}
  \hline\hline
  Coupled channels      &  $\slashed{X}$-UChPT~\cite{Altenbuchinger:2013vwa}         & $X$-UChPT~\cite{Albaladejo:2016eps}        & Coupled channels   & $\slashed{X}$-UChPT~\cite{Altenbuchinger:2013vwa}         & $X$-UChPT~\cite{Albaladejo:2016eps} \\ \hline

  $B\bar{K}$       & $0.020-0.231i$  & $-0.194-0.014i $  & $DK$       & $0.056-0.158i$  & $-0.253-0.038i$\\
  $B_s\pi$   & $-0.006$         & $0.206$         & $D_s\pi$   & $0.004$          & $0.126$ \\
  \hline\hline
\end{tabular}
\end{table}

The scattering length of channel $i$ is defined as
\begin{equation}
a_{ii}=-\frac{1}{8\pi(M_i+m_i)}T_{ii}(s=(M_i+m_i)^2).
\end{equation}
Using the $G$ function determined in Ref.~\cite{Altenbuchinger:2013vwa} and Ref.~\cite{Albaladejo:2016eps},\footnote{
The loop function is always regularized in the dimensional regularization scheme, unless otherwise specified.} we obtain
the scattering lengths of $B\bar{K}$ and $B_s\pi$ tabulated in Table~\ref{Table:twochannel}. Clearly, the results obtained in the two  approaches are quite different.
The scattering lengths, particularly that of $B_s\pi$ obtained in $\slashed{X}$-UChPT~\cite{Altenbuchinger:2013vwa}, show clearly that the interactions are rather weak in the $B_s\pi$ and $B\bar{K}$ coupled channels. This implies that there is no bound state or resonant state, consistent with a direct search on the second Riemann sheet.
For the sake of comparison, Table~\ref{Table:twochannel} also lists the scattering lengths of $DK$ and $D_s\pi$. We note that the scattering lengths obtained in $X$-UChPT~\cite{Albaladejo:2016eps}
are  much larger than those of $\slashed{X}$-UChPT~\cite{Altenbuchinger:2013vwa}, inconsistent with the lQCD results of Ref.~\cite{Liu:2012zya}.

lQCD simulations allow one to understand  more physical observables by studying their quark mass dependence.  In this regard, it is useful to study the $m_{\pi}$ dependence of the scattering lengths.
For such a purpose in the UChPT, one needs the pion mass dependence of the
constituent hadrons, namely, $m_K$, $m_B$ and $m_{B_s}$.
Following  Ref.~\cite{Altenbuchinger:2013vwa}, we  take
\begin{equation}\label{mpidependence}
  \begin{split}
    m_K^2 =& \hat{a} + \hat{b}m_{\pi}^2,\\
    m_B^2 =& m_0^2 + 4c_0(m_{\pi}^2+m_{K}^2) - 4c_1m_{\pi}^2, \\
    m_{B_s}^2 =& m_0^2 + 4c_0(m_{\pi}^2+m_{K}^2) + 4c_1(m_{\pi}^2 -2m_K^2),
  \end{split}
\end{equation}
with $\hat{a}=0.317$, $\hat{b}=0.487$, $c_0=0.015$, and $c_1=-0.513$.  These LECs are fixed using
the experimental data and the lattice QCD masses of Ref.~\cite{Liu:2012zya} as explained above.

\begin{figure}
  \centering
  \includegraphics[width=0.48\textwidth]{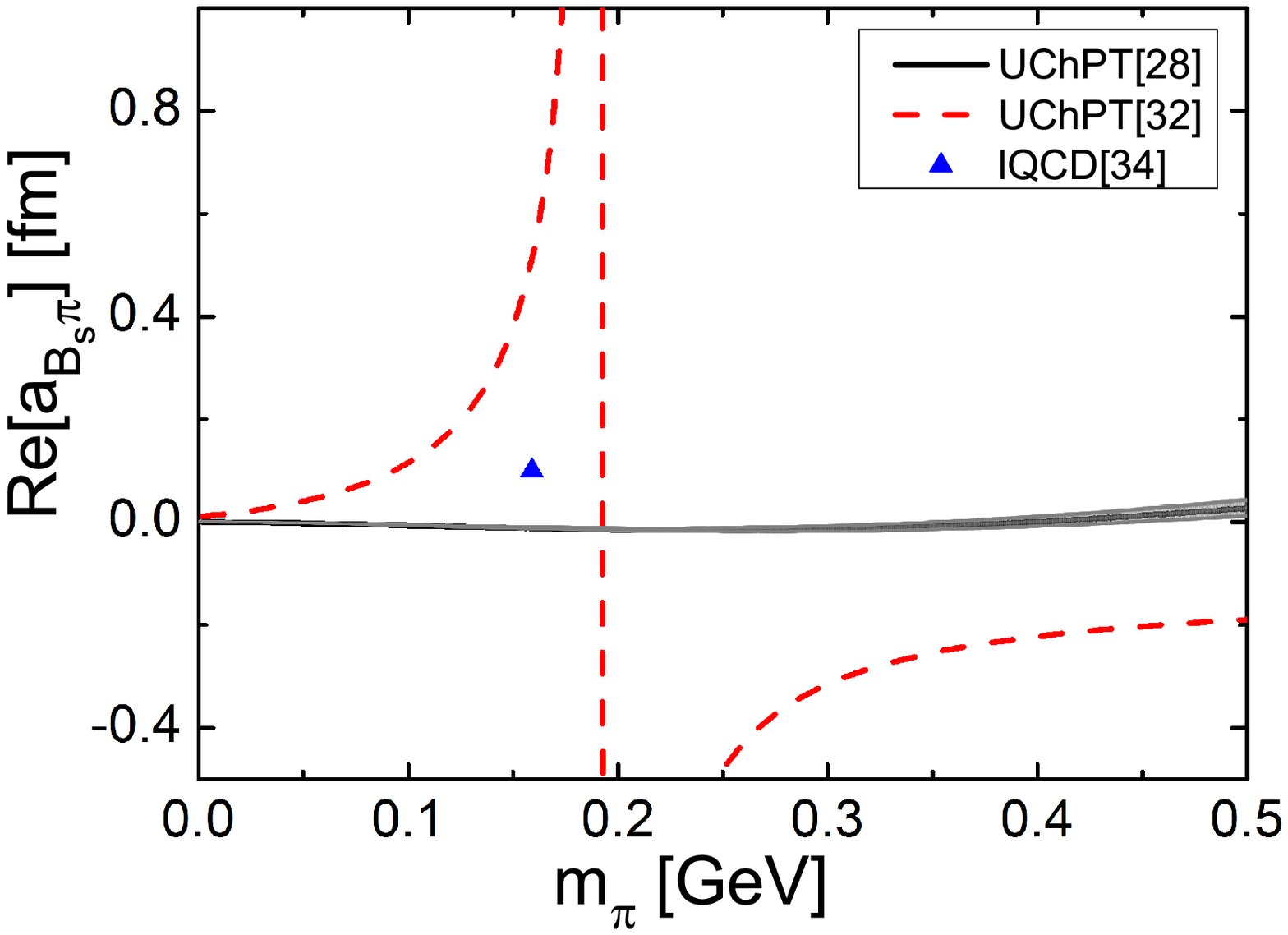}
      \includegraphics[width=0.48\textwidth]{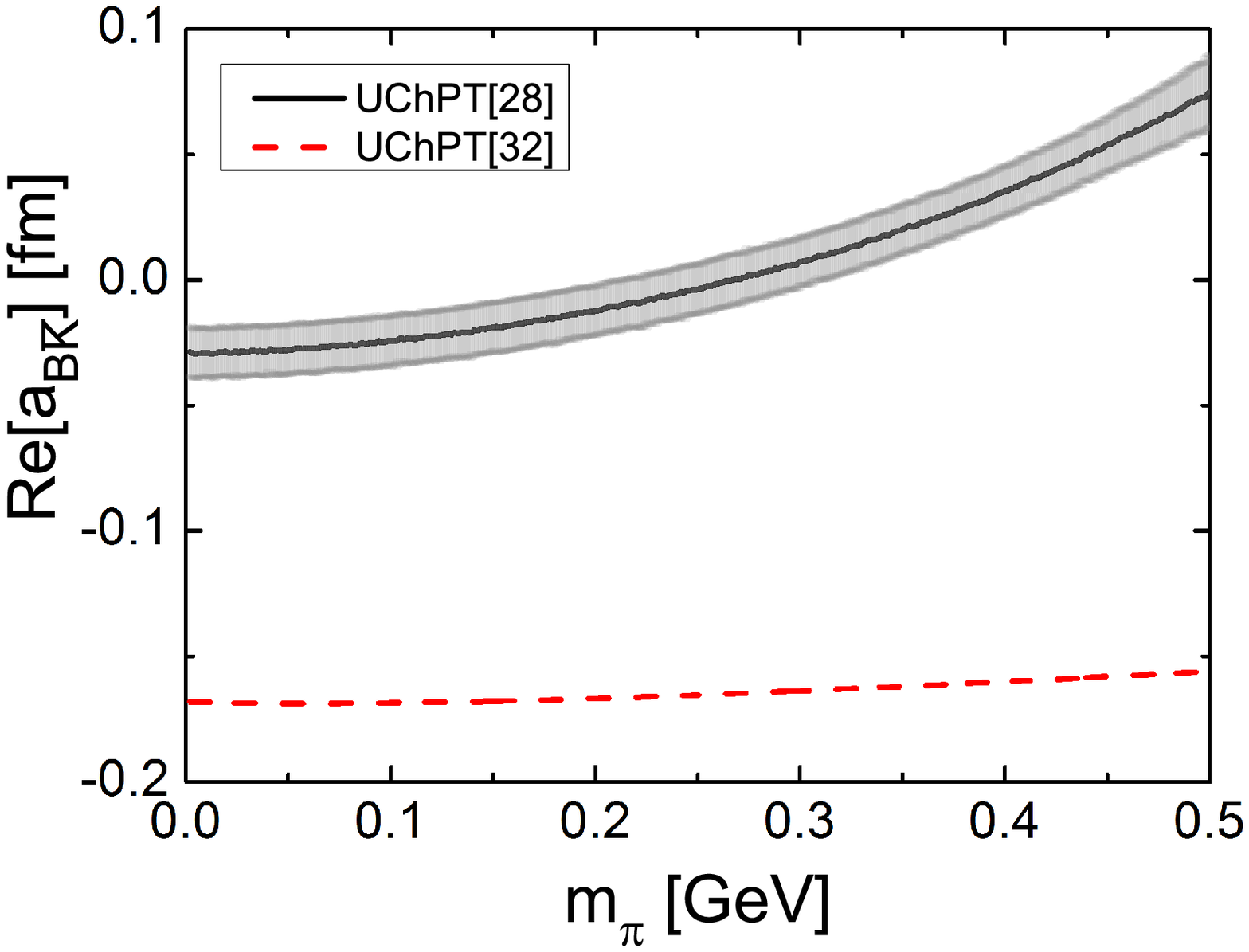}\\
  \caption{ Real part of the scattering lengths $a_{B_s\pi}$ and $a_{B\bar{K}}$ as a function of the pion mass $m_{\pi}$ obtained
  in the $\slashed{X}$-UChPT~\cite{Altenbuchinger:2013vwa} and $X$-UChPT~\cite{Albaladejo:2016eps}. The uptriangle
  in the left panel denotes the lQCD result of Ref.~\cite{Lang:2016jpk}, which is an average of the six data obtained using different sets
  of gauge configurations (see the bottom panel of Fig.~2 of  Ref.~\cite{Lang:2016jpk}). The shaded area indicates uncertainties originating from
  the lQCD data of  Ref.~\cite{Liu:2012zya}.
  \label{ReaBspi}}
\end{figure}

In Fig.~\ref{ReaBspi}, we show the scattering lengths $a_{B_s \pi}$ and $a_{B\bar{K}}$ as a function of the pion mass.
One can see that $a_{B_s\pi}$ shows some ``threshold'' effects.  These effects can easily be understood from the amplitude expressed in terms of couplings and pole positions,
\begin{equation}
T_{ii}(s)=\frac{g_i^2}{\sqrt{s}-\sqrt{s}_0},
\end{equation}
where $g_i$ is the coupling defined in Eq.~(\ref{gi}), and $\sqrt{s}_0$ the pole position.
To calculate scattering lengths, $\sqrt{s}=m_{B_s} +m_\pi$. Once the trajectory of the threshold crosses that of the pole as $m_{\pi}$ varies, a singularity will emerge. In fact, similar effects have already been observed in Ref.~\cite{Zhou:2014ila}. On the other hand, as $m_\pi$ increases, $a_{B\bar{K}}$ stays more or less constant
in $X$-UChPT~\cite{Albaladejo:2016eps}, while it increases slightly in $\slashed{X}$-UChPT~\cite{Altenbuchinger:2013vwa}.
For the sake of comparison, we also show the lQCD result of Ref.~\cite{Lang:2016jpk}.  We note that the lQCD result, obtained with a $\pi$ mass close to its physical value,
lies in between the results of Ref.~\cite{Altenbuchinger:2013vwa,Albaladejo:2016eps}, and thus cannot distinguish the two scenarios.  For such a purpose,  lQCD simulations with quark masses larger than their physical values will be more
useful.

\subsection{Compositeness of $X(5568)$}
The large cutoff used in $X$-UChPT~\cite{Albaladejo:2016eps} indicates that the dynamically generated $X(5568)$ should
contain rather large non-$B_s\pi$ and $B\bar{K}$ components. This can be quantified using
the Weinberg composition condition and its extensions~\cite{Weinberg:1962hj,Weinberg:1965zz,Hanhart:2010wh,Baru:2003qq,Cleven:2011gp,Gamermann:2009uq,YamagataSekihara:2010pj, Aceti:2012dd,Xiao:2012vv,Aceti:2014ala,Aceti:2014wka,Hyodo:2011qc,Hyodo:2013nka,Sekihara:2014kya,Nagahiro:2014mba,Garcia-Recio:2015jsa,Guo:2015daa}.

Following Ref.~\cite{ Aceti:2014ala} we define the weight of a hadron--hadron component in a composite particle
 by
 \begin{equation}
 X_i=-\mathrm{Re}\left[g_i^2\left[\frac{\partial G_i^\mathrm{II}(s)}{\partial \sqrt{s}}\right]_{\sqrt{s}=\sqrt{s_0}}\right],
 \end{equation}
 where $\sqrt{s_0}$ is the pole position, $G_i^\mathrm{II}$ is the loop function evaluated on the second Riemann sheet, and
 $g_i$ is the  coupling of the respective resonance or bound state to channel $i$ calculated  as
 \begin{equation}\label{gi}
 g_i^2=\mathop{\mathrm{lim}}_{\sqrt{s}\rightarrow\sqrt{s_0}}(\sqrt{s}-\sqrt{s_0})T_{ii}^\mathrm{II},
 \end{equation}
 where $T_{ii}^\mathrm{II}$ is the $ii$ element of the $T$ amplitude on the second Riemann sheet.

 The deviation of the sum of $X_i$ from unity is related to the energy dependence of the $s$-wave  potential,
 \begin{equation}
 \sum_i X_i=1-Z,
 \end{equation}
 where
 \begin{equation}
 Z=-\sum_{ij}\left[g_i G_i^\mathrm{II}(\sqrt{s})\frac{\partial V_{ij}(\sqrt{s})}{\partial \sqrt{s}}G_j^\mathrm{II}(\sqrt{s})g_j\right]_{\sqrt{s}=\sqrt{s}_0}. \\
 \end{equation}
The quantity $Z$ is often attributed to the weight of missing channels.

Using $X$-UChPT~\cite{Albaladejo:2016eps}, we  obtain $X_{B\bar{K}}=0.10-0.02i$, $X_{B_s\pi}=0.06+0.11i$,  and $Z=0.83-0.09i$.
The value of $Z$ is much larger than the typical size  for a state dominated by  molecular components, which indicates the missing of contributions of other components. Such a result is consistent with the unusual size of $\Lambda$ and similar conclusions have been drawn in Ref.~\cite{Albaladejo:2016eps}.

One should note that the above defined compositeness and the so-drawn conclusion
are model dependent, which is different from  the original Weinberg criterion.
In this formalism, the compositeness and elementariness are  defined as the fractions of the contributions from the two-body scattering states and one-body bare states
to the normalization of the total wave function within the particular model space, respectively (see, e.g., Ref.~\cite{Sekihara:2014kya}).  In the present case, the small compositeness $1-Z$ simply indicates that in the model space of the $\slashed{X}$-UChPT~\cite{Albaladejo:2016eps},
the meson--meson components only account for a small fraction of the total wave function, and thus $X(5568)$ cannot be categorized as a meson--meson molecule.

\section{$B_s\pi$ and $B\bar{K}$ interactions in finite volume}
If $X(5568)$ exists, one should be able to observe it in a lQCD simulation, which can be anticipated in the near future, given the fact that
the LHCb result has cast doubts on the existence of the $X(5568)$. In view of such possibilities, in the following,
we predict the discrete energy levels that one would obtain in
a lattice QCD simulation. Such an exercise provides a highly non-trivial test of $\slashed{X}$-UChPT~\cite{Altenbuchinger:2013vwa} and $X$-UChPT~\cite{Albaladejo:2016eps}.

In this work, we follow the method proposed in Ref.~\cite{Geng:2015yta} to calculate the loop function $G$ in finite volume in the dimensional regularization scheme. Introducing a finite-volume correction, $\delta G$, $\tilde{G}$ can be written as
\begin{equation}\label{FVC1}
   \tilde{G} = G^D + \delta G,
\end{equation}
where $G^D$ is the loop function calculated in the dimensional regularization scheme, either $G_{\mathrm{HQS}}$ or $G_{\overline{\mathrm{MS}}}$,  and $\delta G$ has the following form~\cite{Geng:2015yta}:
\begin{equation}\label{FVCdG}
  \delta G = -\frac{1}{4} \int_0^1 \mathrm{d}x\delta_{3/2}(\mathcal {M}^2(s)),
\end{equation}
where
\begin{equation}\label{FVCM}
  \mathcal{M}^2(s)=(x^2-x)s+xM^2+(1-x)m^2-i\epsilon.
\end{equation}

For the case of $\sqrt{s}>M+m$, $\delta_{r}(\mathcal{M}^2(s))$ can be written as a sum of the following three parts~\cite{Bernard:2007cm,Ren:2013oaa}:
\begin{equation}\label{FVCM2}
  \delta_{r}(\mathcal{M}^2(s))=g_1^r-g_2^r+g_3^r,
\end{equation}
where  $g_{1,2,3}^r$ are
\begin{equation}\label{FVCM3}
  \begin{split}
    g_1^r=& \frac{1}{L^3} \sum_{\vec{q}} \{ \frac{1}{[\frac{4\pi^2\vec{n}^2}{L^2}+\mathcal{M}^2(s)]^r} -\frac{1}{[\frac{4\pi^2\vec{n}^2}{L^2}+\mathcal{M}^2(m_{ss}^2)]^r} +\frac{r(x^2-x)(s-m_{ss}^2)}{[\frac{4\pi^2\vec{n}^2}{L^2}+\mathcal{M}^2(m_{ss}^2)]^{r+1}} \} ,\\
    g_2^r=& \int_0^{+\infty}\frac{q^2dq}{2\pi^2} \{\frac{1}{[\vec{q}^2+\mathcal{M}^2(s)]^r} -\frac{1}{[\vec{q}^2+\mathcal{M}^2(m_{ss}^2)]^r} +\frac{r(x^2-x)(s-m_{ss}^2)}{[\vec{q}^2+\mathcal{M}^2(m_{ss}^2)]^{r+1}} \},\\
    g_3^r=& \delta_{r}(\mathcal{M}^2(m_{ss}^2)) -r(x^2-x)(s-m_{ss}^2)\delta_{r+1}(\mathcal{M}^2(m_{ss}^2)),
  \end{split}
  \end{equation}
  and $L$ is the spatial size of the lattice.~\footnote{Throughout this paper, we assume a periodic boundary condition for the lQCD setup and that
  the temporal size is much larger than the spatial size and therefore can be taken as infinity.}
The separation scale $m_{ss}$ needs to satisfy $m_{ss}<M+m$. In the case of $\sqrt{s}<M+m$, $\delta_{r}(\mathcal{M}^2(s))$ can be expressed as~\cite{Geng:2011wq}
\begin{equation}\label{FVCM4}
  \delta_{r}(\mathcal{M}^2(s))=\frac{2^{-1/2-r}(\sqrt{\mathcal{M}})^{3-2r}}{\pi^{3/2}\Gamma(r)} \sum_{\vec{n}\neq 0}(L\sqrt{\mathcal{M}^2}|\vec{n}|)^{-3/2+r}K_{3/2-r}(L\sqrt{\mathcal{M}^2}|\vec{n}|),
\end{equation}
where $K_n(z)$ is the modified Bessel function of the second kind, and
\begin{equation}\label{FVCM5}
  \sum_{\vec{n}\neq0}\equiv \sum_{n_x=-\infty}^{\infty}\sum_{n_y=-\infty}^{\infty}\sum_{n_z=-\infty}^{\infty}(1-\delta(|\vec{n}|,0)),
\end{equation}
with $\vec{n}=(n_x,n_y,n_z)$. It should be mentioned that in actual calculations the
discrete summations in Eqs.~ (\ref{FVCM3}--\ref{FVCM5}) are only taken up to a certain number, $|n|_\mathrm{max}= L/(2a)$ with $a$  the lattice spacing.

The Bethe--Salpeter equation in finite volume reads
\begin{equation}\label{BSfv}
\tilde{T}=\frac{1}{V^{-1}-\tilde{G}}.
\end{equation}
The discrete energy levels one would observe in a lattice QCD simulation are determined
via $\det(V^{-1}-\tilde{G})=0$.

\begin{figure}
  \centering
  \includegraphics[width=0.48\textwidth]{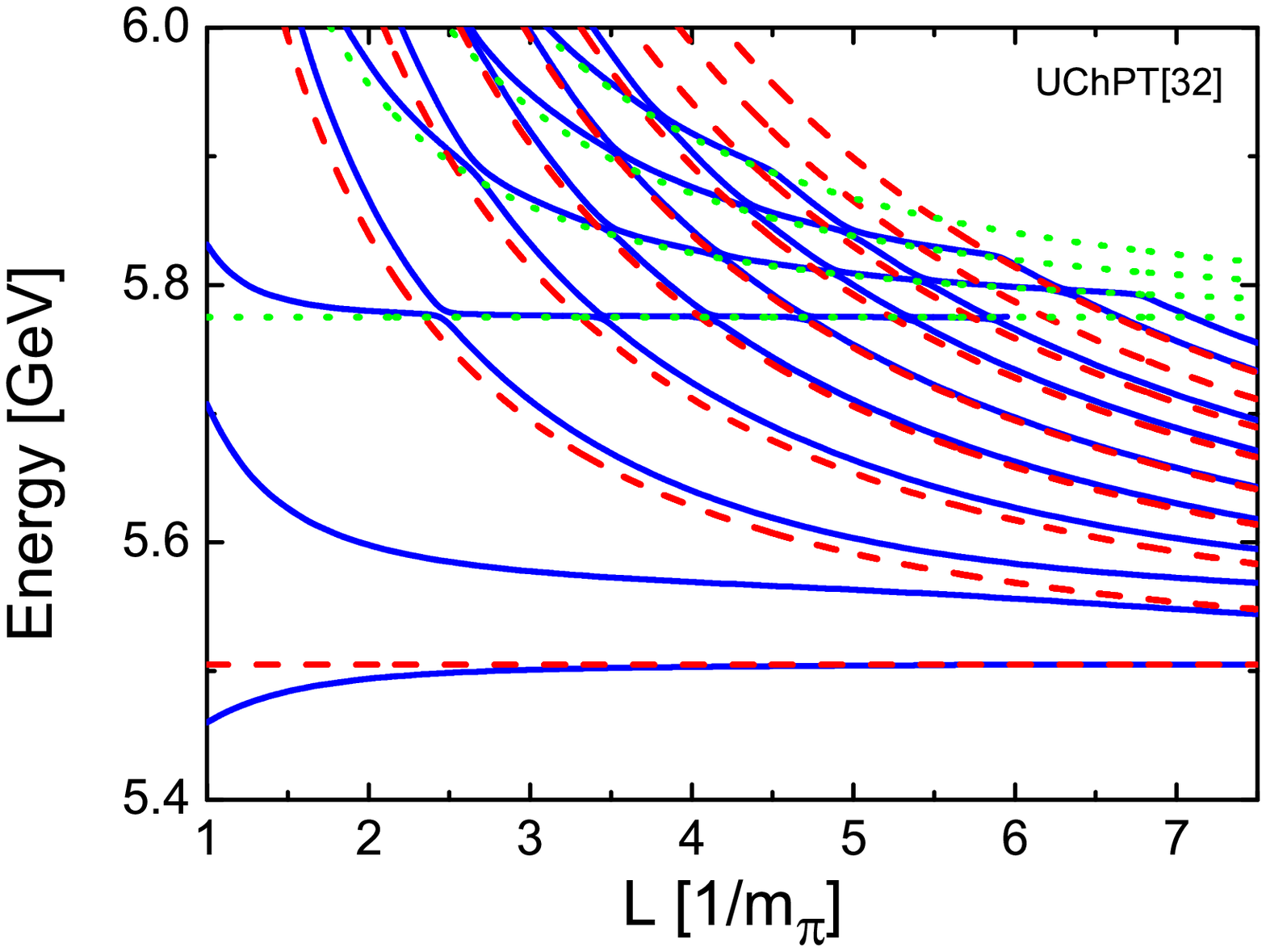}
    \includegraphics[width=0.48\textwidth]{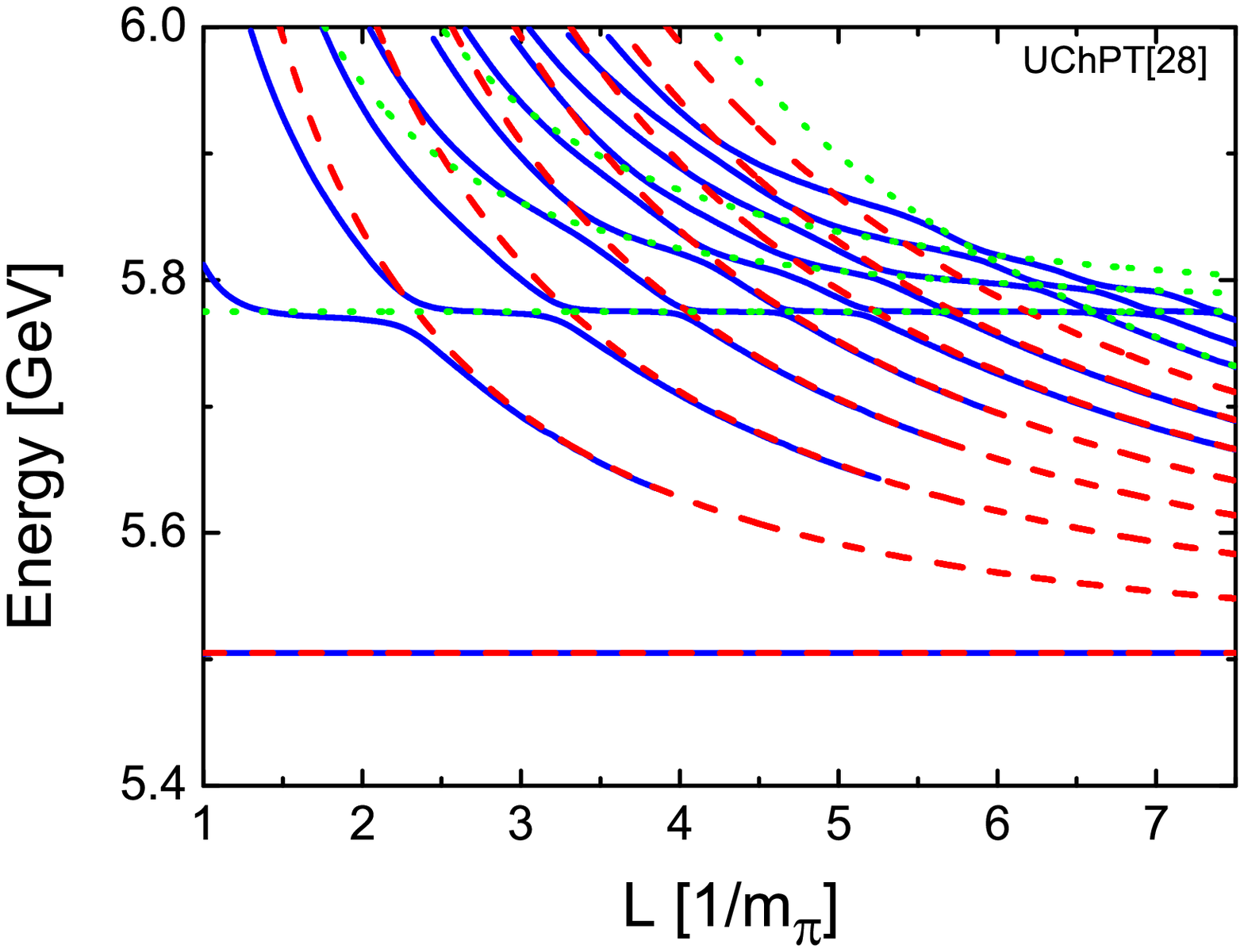}
  \caption{ Discrete energy levels of the $B_s\pi$--$B\bar{K}$ system as a function of the lattice size $L$. Left panel: obtained with
  $X$-UChPT~\cite{Albaladejo:2016eps}; right panel: obtained with $\slashed{X}$-UChPT~\cite{Altenbuchinger:2013vwa}.  The solid lines are the energy levels obtained
  by solving Eq.~(\ref{BSfv}), while the dashed and dotted lines are the energy levels of non-interacting $B_s\pi$ and $B\bar{K}$ pairs, respectively. }\label{Elevel-con}
\end{figure}

\begin{figure}
  \centering
  \includegraphics[width=0.6\textwidth]{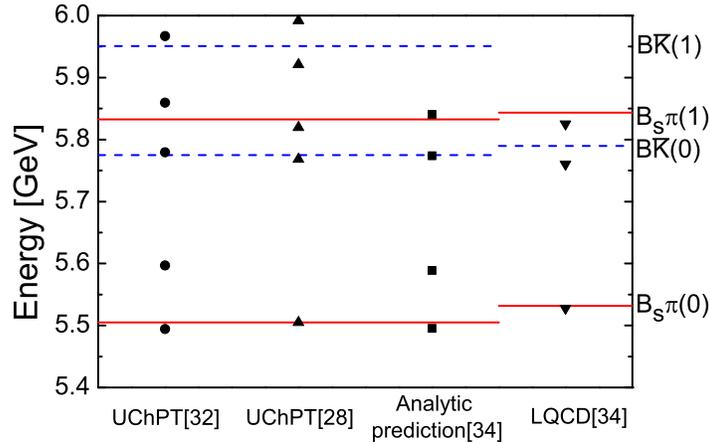}
  \caption{ Discrete energy levels at $L=2.9$ fm obtained in different approaches: the solid points, up triangles, squares, and down triangles correspond to the results of $X$-UChPT~\cite{Albaladejo:2016eps},
  $\slashed{X}$-UChPT~\cite{Altenbuchinger:2013vwa}, the analytic prediction using the L\"uscher method~\cite{Lang:2016jpk}, and the lQCD results
  of Ref.~\cite{Lang:2016jpk}, respectively. The solid (dashed) lines refer to the energy levels of non-interacting $B_s\pi$ ($B\bar{K}$) pairs.} \label{lQCD}
\end{figure}

In Fig.~\ref{Elevel-con}, we show the so-obtained discrete energy levels in both scenarios, $X$-UChPT~\cite{Albaladejo:2016eps} and $\slashed{X}$-UChPT~\cite{Altenbuchinger:2013vwa}.  From the left panel, one can clearly identify an extra energy level, namely, the second energy level, which
can be associated to $X(5568)$. All the other discrete energy levels lie close to one of the free energy levels, $B_s\pi(\ell)$ or $B\bar{K}(\ell)$, where $\ell$ denotes the
energy of the corresponding discrete energy level with energy $E[B_s\pi(\ell)]=\sqrt{m_{B_s}^2+k^2}+\sqrt{m_\pi^2+k^2}$ wth $k=\ell \frac{2\pi}{L}$ $(\ell=0,1,2,\ldots)$ and likewise for
$E[B\bar{K}(\ell)]$.  On the other hand, no extra energy level appears in the right panel, consistent with the fact that the interactions are weak and no resonance or
bound state is found in $\slashed{X}$-UChPT~\cite{Altenbuchinger:2013vwa}.

In a recent study~\cite{Lang:2016jpk}, a lQCD simulation employing the PACS-CS gauge configurations was performed~\cite{Aoki:2008sm}.
It was shown that  no state corresponding to  $X(5568)$ exists in the simulation, consistent with the LHCb result~\cite{lhcb}. In addition, the authors of Ref.~\cite{Lang:2016jpk} provided an analytic prediction based on the L\"uscher method. They included the $X(5568)$ explicitly via a resonant Breit--Wigner-type phase shift and then related the phase shift with the discrete energy levels through the L\"uscher method. For more details we refer to subsection II.A of Ref.~\cite{Lang:2016jpk}. In Fig.~\ref{lQCD}, we compare the lQCD discrete energy levels with those obtained in our present study. Since the quark masses in the lQCD simulation are not yet physical, the non-interacting energy levels of the lQCD are slightly shifted upward compared to those calculated theoretically using
physical meson masses. Apparently, the lQCD results are consistent with $\slashed{X}$-UChPT~\cite{Altenbuchinger:2013vwa}, but
not $X$-UChPT~\cite{Albaladejo:2016eps} in which a large cutoff was used to reproduce the D0 data.  On the other hand,
the analytic predictions based on the  L\"uscher method~\cite{Lang:2016jpk} are
consistent with $X$-UChPT~\cite{Albaladejo:2016eps}, as they should be since in Ref.~\cite{Lang:2016jpk}, the D0 $X(5568)$ mass and width were employed in the L\"uscher method
as inputs and in Ref.~\cite{Albaladejo:2016eps} the only parameter in $X$-UChPT, the subtraction constant, was fixed by fitting to the D0 data.  We should note that for a single channel, the L\"uscher method is consistent with the J\"ulich--Valencia approach adopted in the present work (see, e.g., Ref.~\cite{Geng:2015yta} for an explicit comparison in the case of the $KK^*$ scattering).

From the above comparison, one can conclude that  there is indeed a tension among the D0  data, the lQCD results of Ref.~\cite{Lang:2016jpk},
and indirectly those of Ref.~\cite{Liu:2012zya}, provided that  heavy-quark symmetry and chiral symmetry are not somehow strongly broken.

\section{Summary}
The recent D0 claim of the existence of  $X(5568)$ has aroused a lot of interest. In the present paper, we showed explicitly the tension between the D0
discovery and the lQCD results on the charmed meson--pseduoscalar meson scattering, including the scattering lengths and the existence of $D_{s0}^*(2317)$, from the perspective of approximate chiral symmetry and heavy-quark  symmetry as implemented in the unitary chiral perturbation theory.
We then formulated the unitary chiral description of the coupled channel $B_s\pi$--$B\bar{K}$ interactions in finite volume.  Our results, when compared with the latest lattice QCD simulation, confirm the inconsistency and disfavor the existence of $X(5568)$ in unitary chiral perturbation theory. We conclude that more experimental and theoretical efforts are needed to clarify the current situation.

\acknowledgements
We thank Eulogio Oset for a careful reading of the first version of this manuscript and for his valuable comments.
This work is partly supported by the National Natural Science Foundation of China under Grants  Nos. 11375024, 11522539, 11335002, and 11411130147, the Fundamental Research Funds for the Central Universities, the
Research Fund for the Doctoral Program of Higher Education under Grant No. 20110001110087, and the China Postdoctoral Science Foundation under Grant No. 2016M600845.

\end{document}